\documentclass{article}
\usepackage{amsmath,natbib,graphicx,siunitx}
\usepackage{bigfoot} 
\usepackage[left=1in,top=1in,right=1in,bottom=1in,nohead,paperwidth=8.5in, paperheight=11in]{geometry} 
\title{Lasso-based forecast combinations for forecasting realized variances}
\title{\bf Lasso-based forecast combinations for forecasting realized variances}
\author{Ines Wilms$^a$\footnote{Corresponding author: ines.wilms@kuleuven.be}, Jeroen Rombouts$^b$, Christophe Croux$^a$ 
\\ \textit{\small $^{a}$ Faculty of Economics and Business, KU Leuven}
\\ \textit{\small $^{b}$ ESSEC Business School}
}

\date{ }
\begin{document}
\maketitle

\paragraph{Abstract.}
Volatility forecasts are key inputs in financial analysis. While lasso based forecasts have shown to perform well in many applications, their use to obtain volatility forecasts has not yet received much attention in the literature.
Lasso estimators produce parsimonious forecast models. 
Our forecast combination approach hedges against the risk of selecting a wrong degree of model parsimony. 
Apart from the standard lasso, we consider several lasso extensions that account for the dynamic nature of the forecast model. 
We apply forecast combined lasso estimators in a comprehensive forecasting exercise using realized variance time series of ten major international stock market indices. 
We find the lasso extended ``ordered lasso" to give the most accurate realized variance forecasts.
Multivariate forecast models, accounting for volatility spillovers between different stock markets, outperform univariate forecast models for longer forecast horizons.   \\


\newpage

\section{Introduction}
Volatility forecasts of financial assets are of great importance in finance.  
For instance, risk management, asset allocation, option pricing and trading in financial markets (see e.g. \citealp{Knight07})  require reliable volatility forecasts. 
We use lasso-based  forecast methods \citep{Tibshirani96} to obtain such volatility forecasts of various international stock market indices.  The lasso has become a popular forecast method since it provides parsimonious forecast models (see e.g. \citealp{Fan2011}). Its usefulness for volatility forecasting, however, is underexplored. Our aim is  to investigate
(i)   which lasso-based forecast method provides the most accurate volatility forecasts and, 
(ii)  whether these volatility forecasts become more accurate if spillover effects between the different stock market indices are accounted for, and 
(iii) whether forecast combination in the lasso-based forecast methods improves further the forecast accuracy.



The inherent problem in obtaining volatility forecasts is that volatility is unobservable. 
Therefore, one traditionally considered latent variable models such as stochastic volatility models or GARCH models \citep{Bollerslev86}. 
However, with the advent of high-frequency data, accurate estimates of volatility have become available \citep{Andersen01, Barndorff02}. Realized variances are computed as the sum of squared intraday returns for a particular day, thereby making volatility observable. 
The advantage of working with  realized variances is that traditional time series models like univariate AutoRegressive (AR) and multivariate Vector AutoRegressive (VAR) models can be used for forecasting. 

This paper proposes the use of lasso-based estimators for AutoRegressive and Vector AutoRegressive models for log-realized variance of ten international stock market indices. 
The lasso \citep{Tibshirani96} provides sparse model estimates, meaning that many of the lagged autoregressive parameters are estimated as exactly zero. As such, a  parsimonious model is obtained. Such parsimonious models  reduce estimation error and therefore may yield more accurate forecasts.
Recent papers by \cite{Audrino12} and \cite{Audrino15} discuss estimation of the AR model by the lasso to obtain realized variance forecasts. Our paper differs from theirs in that we not only consider lasso estimation of univariate (AR) models but also of multivariate (VAR) models, see e.g. \cite{Callot16} for forecasting large realized covariance matrices. 
Especially when stock markets are going through a period of high uncertainty, like financial crises, their volatility interdependence tends to increase (e.g. \citealp{Diebold15}, Chapter 4). This justifies the use of VAR models over AR models to capture the spillover structure between the different stock market indices' volatilities (i.e. effects from one stock market index volatility on another).
The parsimony advantage of the lasso becomes even more important for estimating such VAR models since they become intractable for large systems of time series, see e.g. \cite{Gelper15}. In fact, a parsimonious spillover structure  is obtained since only the most important spillovers effects between different stock market indices are estimated non-zero. 

While \cite{Audrino12}, \cite{Audrino15} and \cite{Callot16} only consider the use of the standard lasso, we also consider two extensions of the lasso that embed lag selection into the estimation procedure: the hierarchical lasso (e.g. \citealp{Bien13} for regression models, \citealp{Nicholson16} for time series models) and the ordered lasso (e.g. \citealp{Suo15} for univariate models). In addition to the aim of attaining sparse estimates, the hierarchical and ordered lasso aim at attaining estimates with a low maximal lag order. 
Both estimators are very flexible since they allow each time series to have its own lag structure in each equation of the VAR. Most previous studies on multivariate models, by contrast,  require the strong assumption of a single universal lag order.

An important choice that needs to be made for lasso-based estimators concerns the selection of the sparsity parameter that determines the degree of sparsity of the autoregressive estimates (or ``model parsimony"), as measured by the number of non-zero coefficients. Previous studies either use information criteria or cross-validation to select the ``optimal" sparsity parameter. Forecasts are then obtained from the lasso-based estimator with the selected sparsity parameter. We, in contrast, do not select an optimal sparsity parameter but rather combine forecasts obtained with different sparsity parameters in a forecast combination approach. Such forecast combinations are typically more robust to a wrongly chosen sparsity parameter. 

We apply our forecast models to monthly log-realized variance of ten international stock market indices observed over January, 2000 to April, 2016. While most realized variance time series share similar patterns, the in-sample estimation results reveal that their respective optimal lag lengths selected by the different lasso methods can differ substantially. The ordered lasso has best in-sample fit over all markets. Studying the spillovers when estimating multivariate models allows to identify which markets drive others. By measuring  out-of-sample forecast accuracy in terms of absolute forecast errors in an expanding window setup, it turns out that the ordered lasso also dominates.  Forecast combination slightly improves the results, though applying the model confidence set approach shows that the difference with the ordered lasso method with a single sparsity parameter is not significant. Another interesting finding is that the multivariate forecasting models outperform univariate models for the longer forecast horizons we consider (i.e. 3 and 6 months) meaning that spillover effects have long term forecasting power.

The remainder of this paper is structured as follows. 
Section \ref{model_estimators} discusses the univariate and multivariate forecast models, together with the three lasso-based estimators used to obtain log-realized variance forecasts.
Section \ref{results} describes the data and discusses the results on lag length estimation for the univariate forecast model and spillover structure for the multivariate forecast model.
Section \ref{ForecastResults} details the out-of-sample forecast performance of the different forecast methods.
Section \ref{robcheck} provides a range of robustness checks.
Section \ref{conclusion} concludes.

\section{Modeling and forecasting realized variance \label{model_estimators}}
First, we discuss the univariate and multivariate forecast models for the log-realized variances we study in this paper.
Next, we introduce the three lasso-based estimators to obtain estimates and forecasts. Finally, we explain how we perform forecast combination.

\subsection{Modeling  realized variances}
\paragraph{Univariate forecast model}
Denote the log-realized variance of stock market index $i$ at time point $t$ by $\text{log RV}_t^{(i)}$. 
We consider an AutoRegressive (AR) Model of order $p$ for the log-realized variance of stock market index $i$
\begin{equation}
\text{log RV}_t^{(i)} = \beta_1\text{log RV}_{t-1}^{(i)} + \beta_2\text{log RV}_{t-2}^{(i)} + \ldots + \beta_p\text{log RV}_{t-p}^{(i)} + \epsilon_{t}^{(i)}, 
\end{equation}
for $t=p+1,\ldots,T$. The log-realized variance at time point $t$ is modeled as a function of its own past values up to $p$ periods ago.
We assume, without loss of generality, that  all time series are centered such that no intercept is included, and that the error terms $\epsilon_t$ are white noise with mean zero, variance $\sigma^2$.

\paragraph{Multivariate forecast model}
As a multivariate extension, we consider the Vector AutoRegressive (VAR) Model of order $p$ for the log-realized variances of the $q=10$ stock market indices
\begin{equation}
\left\{ \begin{array}{ccl}
\text{log RV}_t^{(1)}&=& \sum_{j=1}^{q}\sum_{l=1}^{p} \beta_{1j,l}\text{log RV}_{t-l}^{(j)} + \epsilon_{t}^{(1)}\\
\text{log RV}_t^{(2)}&=& \sum_{j=1}^{q}\sum_{l=1}^{p} \beta_{2j,l}\text{log RV}_{t-l}^{(j)} +  \epsilon_{t}^{(2)}\\
\vdots & = & \vdots \\
\text{log RV}_t^{(q)}&=& \sum_{j=1}^{q}\sum_{l=1}^{p} \beta_{qj,l}\text{log RV}_{t-l}^{(j)} +  \epsilon_{t}^{(q)},\\
\end{array} \right.
\end{equation}
where the log-realized variance of stock market index $i$ at time point $t$ is modeled as a function of its own past values and the past values of the other stock market indices up to $p$ periods ago. We assume that the multivariate error time series with $q$ components, $(\epsilon_{t}^{(1)},\ldots,\epsilon_{t}^{(q)})'$, is a 
multivariate white noise with mean $\bf 0$, and covariance matrix $\boldsymbol{\Sigma}$. The VAR model, in contrast to the AR model, captures spillovers between the different stock market indices. Indeed $\beta_{ij,l}$ measures the effect at lag $l$ of stock market index $j$ on stock market index $i$. We call this effect a ``spillover" effect if stock market index $j\neq i$.

\subsection{Lasso-based estimators and realized variance forecasts}
We use three lasso-based estimators to obtain realized variance forecasts. 
They differ  in terms of the sparsity structure they impose on the autoregressive parameters.

\paragraph{Lasso}
The lasso estimator with sparsity parameter $\lambda>0$ is given by
\begin{equation} \label{lasso}
\small{
\widehat{\boldsymbol\beta}_{\text{lasso},\lambda} = \underset{\boldsymbol{\beta}}{\operatorname{argmin}} \ \sum_{i=1}^{q}\left(\dfrac{1}{2} \sum_{t=p+1}^{T} \left(\text{log RV}_t^{(i)}- \sum_{j=1}^q \sum_{l=1}^p \beta_{ij,l}\text{log RV}_{t-l}^{(j)}\right)^2 \right) + \lambda \sum_{i=1}^q \sum_{j=1}^q  \sum_{l=1}^p |\beta_{ij,l}|,}
\end{equation}
where the first term in equation \eqref{lasso} is a least squares criterion.
For the univariate forecast model, $q=1$ and the vector $\boldsymbol\beta$ collects all $p$ estimated parameter values. For the multivariate model,  $\boldsymbol\beta$ collects all $q\times q \times p$ estimated parameter values.
The larger the sparsity parameter $\lambda$, the more elements of $\widehat{\boldsymbol\beta}$ will be put to zero, hence the more parsimonious the forecast model.

A drawback of the lasso is that it does not account for the dynamic nature of the AR/VAR model. Indeed, it allows higher order lags of a particular time series to be included in the model while lower order lags of the same time series are absent. We might want a higher order lag  to be included in the model only if its lower order lags are already included. To accommodate this feature, we consider two extensions of the lasso.

\paragraph{Hierarchical Lasso}
The hierarchical lasso estimator \citep{Bien13, Nicholson16} with sparsity parameter $\lambda>0$ is given by
\begin{equation} \label{hierlasso}
\small{
\widehat{\boldsymbol\beta}_{\text{hierlasso},\lambda} = \underset{\boldsymbol{\beta}}{\operatorname{argmin}} \ \sum_{i=1}^{q}\left(\dfrac{1}{2} \sum_{t=p+1}^{T} \left(\text{log RV}_t^{(i)}- \sum_{j=1}^q \sum_{l=1}^p \beta_{ij,l}\text{log RV}_{t-l}^{(j)}\right)^2 \right) + \lambda \sum_{i=1}^q \sum_{j=1}^q  \sum_{l=1}^p ||\boldsymbol\beta_{ij,(l:p)}||_2,}
\end{equation}
where $||\cdot||_2$ denotes the Euclidean norm of a vector  and $\boldsymbol\beta_{ij,(l:p)}=(\beta_{ij,l},\ldots,\beta_{ij,p})$ is an $p-l+1$ vector collecting the  furthest lags of time series $j$ in equation $i$. The hierarchical lasso accounts for the dynamic nature of the AR/VAR by forcing lower order lagged coefficients of a time series in one of the equations to be selected before its higher order lagged coefficients: for all $l^\prime>l, \widehat{\beta}_{ij,l} = 0$ implies $\widehat{\beta}_{ij,l^\prime} = 0$ almost surely.

\paragraph{Ordered Lasso}
The ordered lasso estimator \citep{Suo15} with sparsity parameter $\lambda>0$ is given by
\begin{equation} \label{orderedlasso}
\small{
\widehat{\boldsymbol\beta}_{\text{orderedlasso},\lambda} = \underset{\boldsymbol{\beta}}{\operatorname{argmin}} \ \sum_{i=1}^{q}\left(\dfrac{1}{2} \sum_{t=p+1}^{T} \left(\text{log RV}_t^{(i)}- \sum_{j=1}^q \sum_{l=1}^p \beta_{ij,l}\text{log RV}_{t-l}^{(j)}\right)^2 \right) + \lambda \sum_{i=1}^q \sum_{j=1}^q  \sum_{l=1}^p (\beta^+_{ij,l} + \beta^-_{ij,l}),}
\end{equation}
subject to the constraints $\beta^+_{ij,1} \geq \beta^+_{ij,2} \geq \ldots \geq \beta^+_{ij,p} \geq 0$, and $\beta^-_{ij,1} \geq \beta^-_{ij,2} \geq \ldots \geq \beta^-_{ij,p} \geq 0$, and each $\beta_{ij,l}=\beta_{ij,l}^+ - \beta_{ij,l}^-$, for all $1\leq i,j \leq q$ and $1\leq l \leq p$.
Similar to the hierarchical lasso, the ordered lasso encourages lower order lagged coefficients to be selected before its higher order lagged coefficients. Additionally, the ordered lasso encourages the absolute values of the lagged effects of each time series $j$ in each equation $i$ to be monotone non-increasing as we move farther back in time.

\paragraph{Automatic lag selection} An advantage of these lasso estimators is that they perform automatic lag selection due to the sparsity induced in $\widehat{\boldsymbol{\beta}}$. In the univariate forecast model, the lag length is estimated as
\begin{equation}
\widehat{p} = \text{sup} \{ 1 \leq l \leq p: \widehat{\beta}_l \neq 0 \}, \nonumber \label{UNIVp}
\end{equation}
the lag corresponding to the largest non-zero estimated parameter. 
Similarly in the multivariate forecast model, the lag length $p_{ij}$ for each time series $1 \leq j \leq q$ in equation $1 \leq i \leq q$ of the VAR is estimated as 
\begin{equation}
\widehat{p}_{ij} = \text{sup} \{1 \leq l \leq p: \widehat{\beta}_{ij,l} \neq 0 \}. \nonumber \label{MULTIp}
\end{equation}
Each time series is allowed to have its own lag structure in each equation of the VAR. 

\paragraph{Algorithmic details} The lasso optimization problems \eqref{lasso}, \eqref{hierlasso}, and \eqref{orderedlasso} can be  solved efficiently using proximal gradient methods. Proximal gradient methods extend traditional gradients methods to non-smooth objective functions (see e.g. \citealp{Bien16}).
Starting from an initial value, the estimate of the parameter vector is updated until converge by evaluating the proximal operator at the gradient step we would take if we were minimizing the least squares criterion alone. 
The key step  is thus to evaluate the proximal operator which involves solving a convex optimization problem.
While the lasso and hierarchical lasso use the Fast Iterative Soft-Thresholding Algorithm to evaluate the proximal operator, the ordered lasso uses the Pool Adjacent Violators Algorithm. For more details, we refer to \cite{Suo15} and \cite{Jenatton11} and references therein. All computations are carried out in \verb|R| version 3.2.1. The code of the algorithms relies on the \verb|R|-packages \verb|orderedLasso| \citep{Rorderedlasso} and \verb|BigVAR| \citep{RBigVAR} and is available from the authors upon request.\\

\subsection{Forecast combination \label{forecastcomb}}
The selection of the sparsity parameter $\lambda$ poses an important choice for lasso estimators. In contrast to previous studies on lasso-based forecasting, we do not select an ``optimal" sparsity parameter. Instead, we combine forecasts obtained from lasso estimations with several sparsity parameters. We obtain several forecasts, one for each choice of the sparsity parameter $\lambda$, and weigh these individual forecasts to construct a final log-realized variance forecast. We expect this forecast combination approach to be more robust to a wrongly chosen sparsity parameter.

To construct our different forecasts, we consider a logarithmic spaced grid of sparsity parameters  of length $L=20$ in the interval $[\lambda_{1}, \lambda_{L}]$, with $\lambda_{1}=0$ and $\lambda_{L}$ an estimate of the sparsity parameter that shrinks all parameters to zero, see \cite{Friedman10} for the lasso. Similar expressions are obtained for the hierarchical and ordered lasso.
If the number of parameters to be estimated is larger than the number of time points, we take $\lambda_{1}=\lambda_{L}/10$.
For each value $\lambda_m$, for $1\leq m \leq L$, we estimate the (AR or VAR) model using either the 
lasso from equation \eqref{lasso}, the hierarchical lasso from equation \eqref{hierlasso}, or the ordered lasso from equation \eqref{orderedlasso}. 
This results in $L$ different estimates $\widehat{\boldsymbol{\beta}}_{\lambda_m}$ having their own degree of sparsity with corresponding $h$-step ahead direct forecasts of the log-realized variance for stock market index $i$  
 \begin{equation}
\widehat{\text{log RV}}_{t+h,\lambda_m}^{(i)} = \sum_{j=1}^q \sum_{l=1}^p (\widehat{\boldsymbol{\beta}}_{\lambda_m})_{ij,l} \text{log RV}_{t+1-l}^{(i)}. \nonumber
 \end{equation}
We use direct forecasts since these are more robust to model misspecification than iterated forecasts \citep{Marcellino06}. 
The final forecast of the log-realized variance of stock market index $i$ at time point $t+h$ is then given by 
\begin{equation}
\widehat{\text{log RV}}_{t+h}^{(i)} =\sum_{m=1}^L w_m \cdot \widehat{\text{log RV}}^{(i)}_{t+h,\lambda_m}, \label{FCeq}
\end{equation}
the weighted sum of the $L$ individual log-realized variance forecasts 
obtained from the lasso, hierarchical or ordered lasso with sparsity parameter $\lambda_m$. As forecast combination weights $w_m$ we take 
\begin{equation}
w_m= \dfrac{\exp{(-0.5\text{BIC}_{\lambda_m})}}{\sum_{m=1}^L \exp{(-0.5\text{BIC}_{\lambda_m})}}. \label{BICweight}
\end{equation} 
Hence, the weights depend on the Bayesian Information Criterion $\text{BIC}_{\lambda_m}$ of each forecast model $1\leq m \leq L$ (e.g. \citealp{Elliott16}, Chapter 14) as given by
\begin{equation}
\text{BIC}_{\lambda_m} = n \cdot\text{log}(\text{Loss}_{\lambda_m}) + \text{df}_{\lambda_m} \cdot \text{log}(n), \nonumber
\end{equation}
where $n=T-p$, $\text{Loss}_{\lambda_m}$ is the first term in objective functions \eqref{lasso}, \eqref{hierlasso}, and \eqref{orderedlasso} using sparsity parameter $\lambda_m$, and  $\text{df}_{\lambda_m}$ is the number of non-zero components of $\widehat{\boldsymbol{\beta}}_{\lambda_m}$. When forecasting the log-realized variances, we take a maximum lag length of $p=36$. The highest weight is given to the model having the smallest value of the BIC. 

Our forecast combination approach is  similar to Bayesian Model Averaging (BMA, see e.g. \citealp{Koop03}, Chapter 11).  BMA gives a final forecast by weighting different models according to their posterior model probability. Hence, BMA is mainly motivated as a way of dealing with model uncertainty.
Note that the lasso can be seen as the mode of a Bayesian estimator with independent double-exponential prior \citep{Tibshirani96}. The sparsity parameter $\lambda$ enters as a hyperparameter in the prior distribution. Hence, we have different models, one for each choice of the hyperparameter $\lambda$ entering the prior. As such, our forecast combination approach deals with uncertainty regarding the appropriate choice of the hyperprameter $\lambda$. Since we use BIC-based weights to construct our final forecast, we also give the highest weight to the model with the highest probability given the data. 

\section{Data and in-sample results \label{results}}
In this section, we discuss the data, 
 the estimated lag lengths of the univariate models,
and  the spillover structure of the multivariate models.

\subsection{Data \label{data}}
We use data on realized variances for ten different stock market indices. We consider a selection of American, Asian and  European stock market indices:
(1) the Amsterdam Exchange Index ``\verb|AEX|", 
(2) the French stock index Cotation Assist\'ee en Continu ``\verb|CAC|", 
(3) the German stock index Deutscher Aktienindex ``\verb|DAX|", 
(4) the American Dow Jones Industrial Average ``\verb|DJIA|", 
(5) the stock index of the Eurozone ``\verb|EUROSTOXX|", 
(6) the United Kingdom Financial Times Stock Exchange Index 100 ``\verb|FTSE|", 
(7) the American stock index ``\verb|NASDAX|", 
(8) the Japanese stock index ``\verb|NIKKEI|", 
(9) the Swiss market index ``\verb|SMI|", and 
(10) the American Standard \& Poor's 500 market index ``\verb|SP500|".

\begin{figure}
 \centering
 \includegraphics[width=13cm]{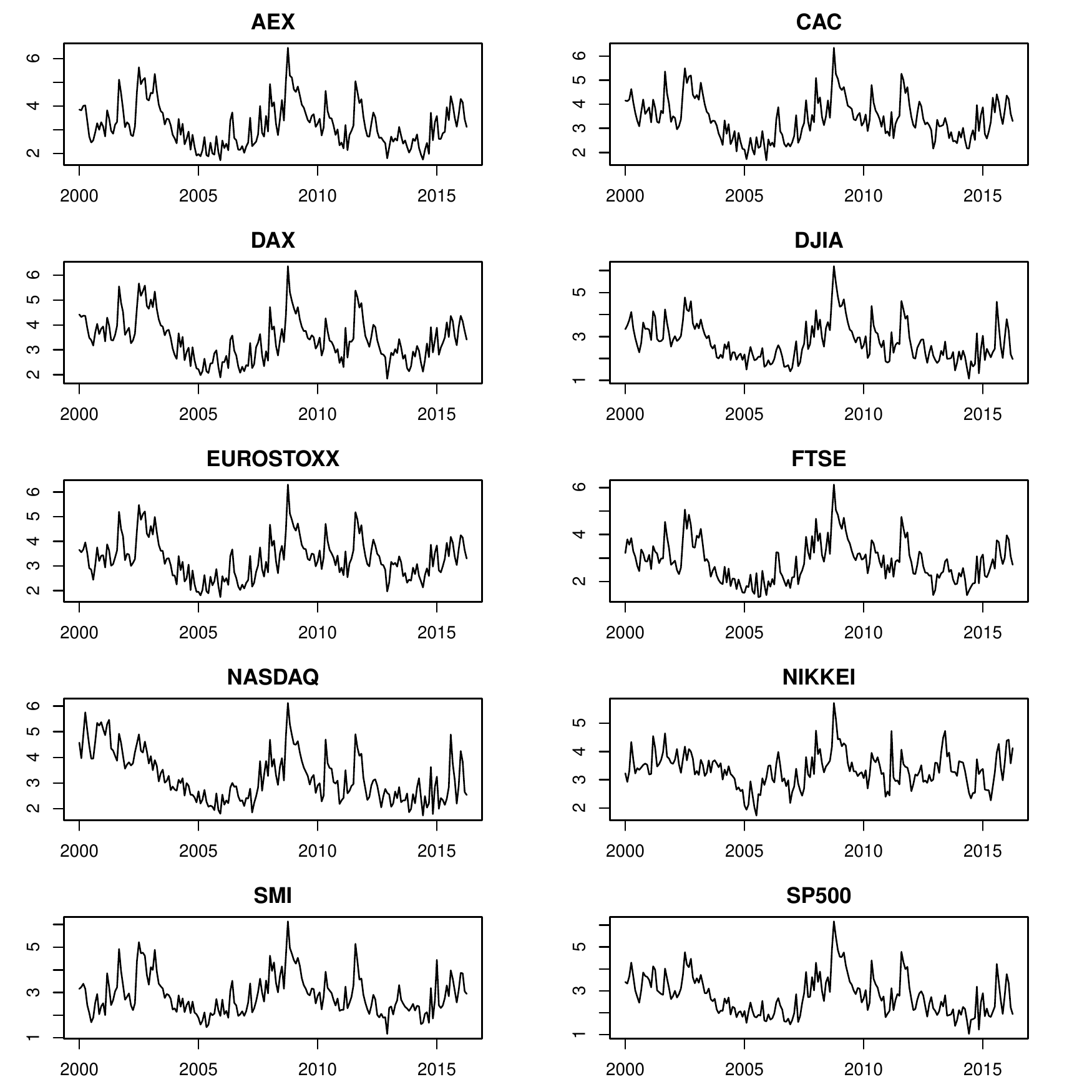}
 \caption{Log-realized variances for the ten stock market indices from January 2000 to April 2016. \label{LogRV}}
 \end{figure}

Daily  realized variance measures based on five minute returns are taken from Oxford-Man Institute of 
Quantitative Finance (publicly available on http://realized.oxford-man.ox.ac.uk/ data/download).
One of our objectives is the comparison of univariate and multivariate forecast models. 
We therefore aggregate the data to monthly frequency between January 2000 and April 2016, i.e. $T=196$ observations.
An advantage of working with monthly data instead of daily data is that we avoid the problem of non-overlapping trading days in the multivariate forecast model of the different international stock market indices.
Note that in Section \ref{conclusion} below, we compare our monthly univariate forecasting results with the forecasts from models estimated on daily realized variances.
Following standard practice,  realized variance is log-transformed ensuring that their distribution more closely resembles the normal distribution and lives on the whole real line. 
Figure \ref{LogRV} plots the log-realized variances. The Augmented Dickey-Fuller test confirms that the time series are stationary. 
Descriptive statistics  are given in Table \ref{Descriptives}. After log-transforming each time series, its skewness and kurtosis more resemble that of a normal distribution.  The log-realized variance of $\verb|NASDAQ|$ is the most volatile and it has the highest first order autocorrelation. The log-realized variance of \verb|NIKKEI|, on the contrary,  is the least volatile and  has the lowest first order autocorrelation.  \\


\renewcommand{\baselinestretch}{1}	
\begin{table}
\caption{Descriptives: Mean, standard deviation, skewness, kurtosis and first order autocorrelation of each log-realized variance time series. \label{Descriptives}}
\centering
\begin{tabular}{lccccc}
  \hline\hline
 & Mean & St.Dev. & Skewness & Kurtosis & $\text{AutoCorr}_{\text{lag=1}}$ \\ 
  \hline
\verb|AEX| & 3.210 & 0.901 & 0.756 & 3.242 & 0.764 \\ 
\verb|CAC| & 3.401 & 0.843 & 0.494 & 3.056 & 0.758 \\ 
\verb|DAX| & 3.434 & 0.870 & 0.598 & 3.102 & 0.780 \\ 
\verb|DJIA| & 2.760 & 0.871 & 0.878 & 3.760 & 0.745 \\ 
\verb|EUROSTOXX| & 3.293 & 0.821 & 0.621 & 3.383 & 0.749 \\ 
\verb|FTSE| & 2.864 & 0.865 & 0.681 & 3.425 & 0.770 \\ 
\verb|NASDAQ| & 3.241 & 0.959 & 0.687 & 2.640 & 0.800 \\ 
\verb|NIKKEI| & 3.361 & 0.608 & 0.357 & 3.945 & 0.659 \\ 
\verb|SMI| & 2.882 & 0.864 & 0.928 & 3.660 & 0.727 \\ 
\verb|SP500| & 2.776 & 0.896 & 0.773 & 3.462 & 0.767 \\ 
   \hline\hline
\end{tabular}
\end{table}
	\renewcommand{\baselinestretch}{1.6}

\subsection{Lag length estimation for univariate  models}
Table \ref{UNIV_lag} reports the estimated lag lengths and the  BIC values of the ten stock market indices for the forecast combined lasso, hierarchical lasso and ordered lasso applied to the full sample. The estimated lag length of each stock market index $i$ is given by $$\widehat{p}_i=\sum_{m=1}^L w_m \widehat{p}_{i,\lambda_m}.$$
In line with our expectations, the hierarchical and ordered lasso give much lower estimated lag lengths than the lasso for the majority of stock market indices (exceptions are \verb|EUROSTOXX|, \verb|NIKKEI|, \verb|SMI| and  \verb|SP500|). 
For the hierarchical and ordered lasso, the lower order lags of the maximum non-zero lag are all non-zero, thereby encouraging a parsimonious, low maximal lag order model. For the lasso, in contrast, the estimated lag length is the highest order non-zero lagged coefficient; lower order lags can be both zero or non-zero. Among the three estimators, the ordered lasso consistently gives the best in-sample fit, i.e. lowest value of BIC for each stock market index.

	\renewcommand{\baselinestretch}{1}
\begin{table}
\caption{Estimated lag lengths $\widehat{p}$ and  BIC (full sample). \label{UNIV_lag}}
\centering
\begin{tabular}{lllccccccccccc}
  \hline\hline
Stock market index &&& \multicolumn{3}{c}{Lasso} && \multicolumn{3}{c}{Hierarchical Lasso} && \multicolumn{3}{c}{Ordered Lasso}\\ 
&&& $\widehat{p}$ && \text{BIC} && $\widehat{p}$ && \text{BIC} && $\widehat{p}$ && \text{BIC} \\
  \hline
\verb|AEX| &&& 4.63 && 173.44 && 5.00 && -69.47 && 3.00 && -160.07 \\ 
\verb|CAC| &&& 31.00 && -99.16 && 5.00 && -84.11 && 2.99 && -174.96 \\ 
\verb|DAX| &&& 4.93 && -140.66 && 4.00 && -96.02 && 2.01 && -181.90 \\ 
\verb|DJIA| &&& 3.45 && -42.32 && 6.02 && -58.83 && 4.92 && -149.00 \\ 
\verb|EUROSTOXX| &&& 2.00 && -112.83 && 6.00 && -86.12 && 2.98 && -178.05 \\ 
\verb|FTSE| &&& 35.09 && -77.24 && 6.01 && -57.51 && 3.00 && -182.95 \\ 
\verb|NASDAQ| &&& 25.16 && -71.52 && 7.00 && -30.88 && 3.97 && -154.92 \\ 
\verb|NIKKEI| &&& 1.00 && -148.62 && 3.85 && -184.02 && 1.45 && -232.36 \\ 
\verb|SMI| &&& 1.00 && 147.33 && 4.00 && -89.18 && 4.53 && -157.50 \\ 
\verb|SP500| &&& 3.34 && 142.56 && 5.00 && -58.54 && 4.66 && -147.03 \\ 
   \hline\hline
\end{tabular}
\end{table}
	\renewcommand{\baselinestretch}{1.6}

The estimated lag lengths of the American stock market indices are typically the highest, followed by the European stock market indices. 
The American stock market indices  have, overall, the highest volatility (see Table \ref{Descriptives}), and contain several volatility spikes. For more turbulent time series like these, higher order AR models typically have more explanatory power than lower order AR models.
By contrast, the estimated lag length for the Japanese stock market index NIKKEI is the lowest.
Since it also has the lowest stock market volatility, it is characterized by calmer market conditions. Simple, low lag order AR models then perform best. \\

\subsection{Spillover structure in multivariate  models}
\begin{figure}
\centering
\includegraphics[width=8cm]{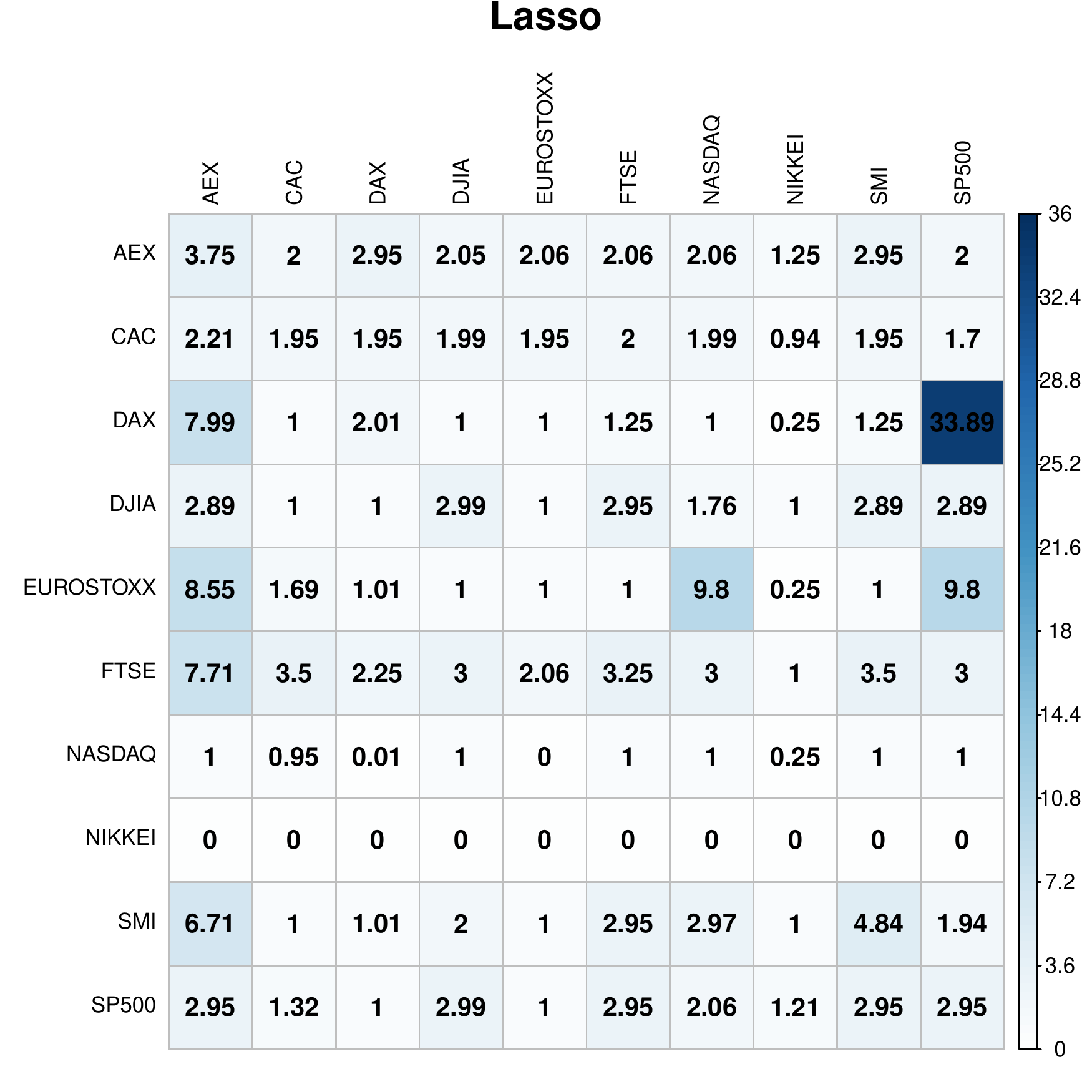}
\includegraphics[width=8cm]{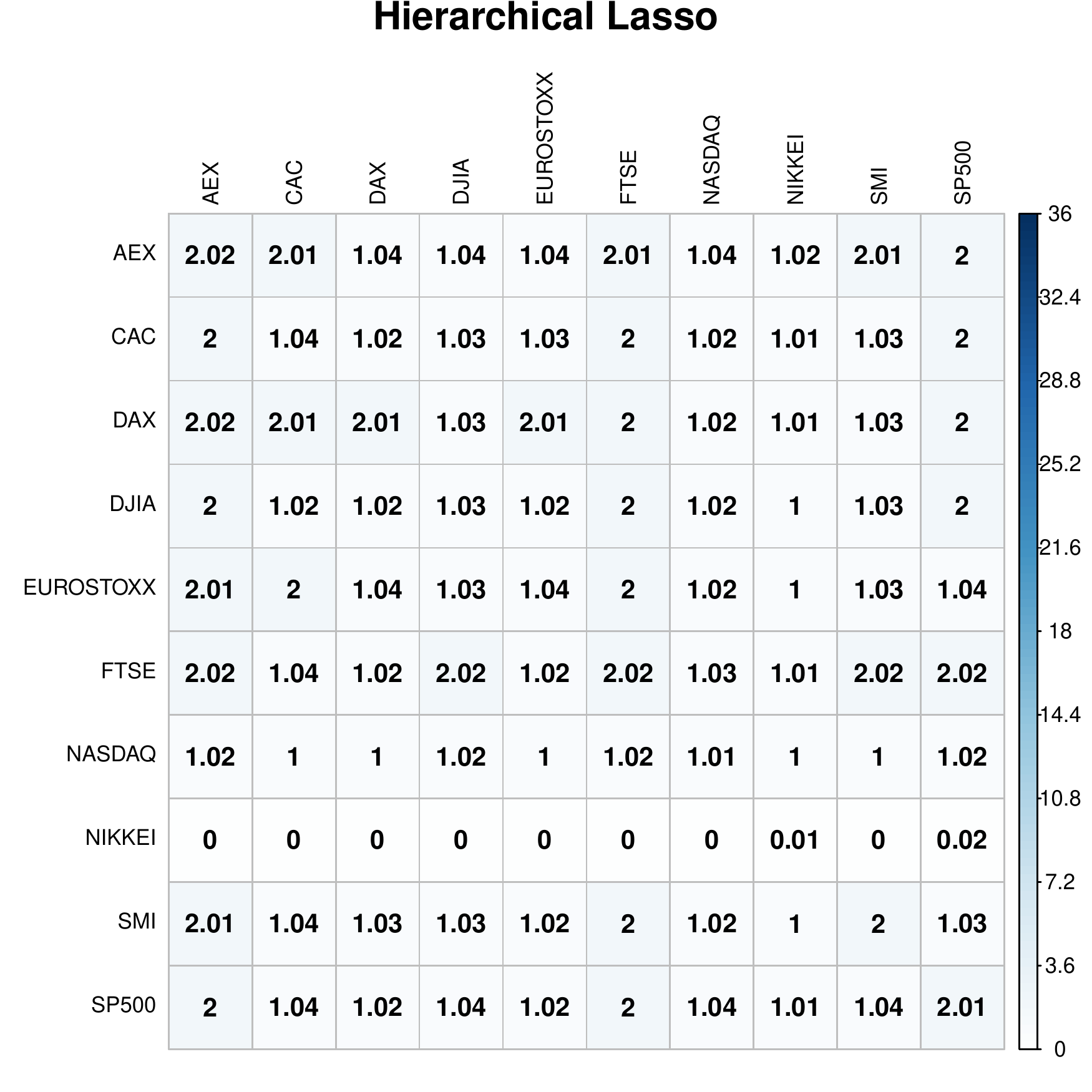} \vspace{1cm}

\includegraphics[width=8cm]{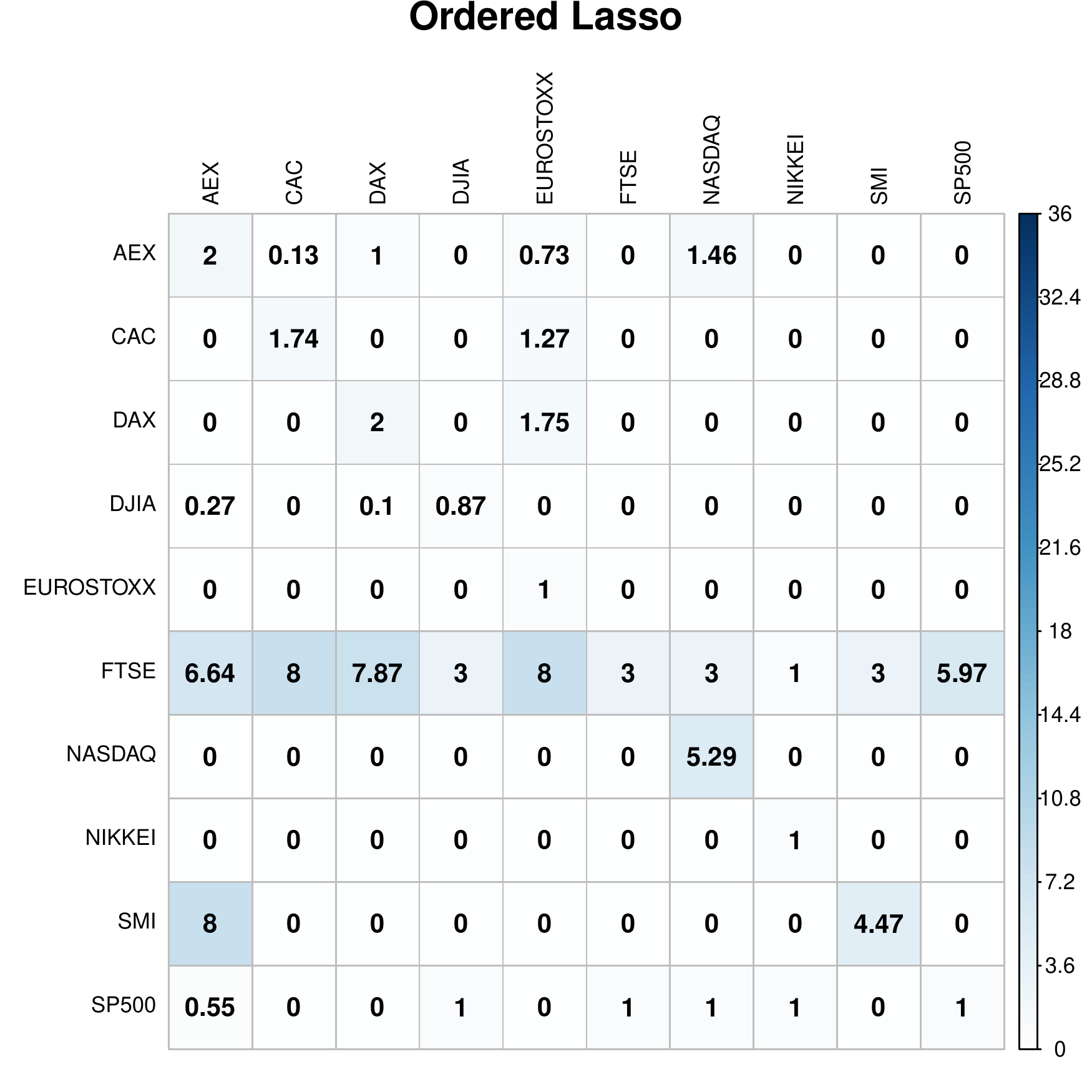}
\caption{Lag matrix of estimated lag lengths. Each cell $ij$ indicates the maximum number of non-zero lags of time series $i$ (row) in equation $j$ (column) of the VAR model estimated by the lasso (top left), hierarchical lasso (top right), or ordered lasso (bottom). The darker the cell, the higher the estimated lag length. \label{Spillover_MULTI}}
\end{figure}

Figure \ref{Spillover_MULTI} visualizes the lag matrix of estimated lag lengths for the forecast combined lasso, hierarchical lasso and ordered lasso applied to the full sample.  For each cell $ij$, the estimated lag length is $$\widehat{p}_{ij}=\sum_{m=1}^L w_m \widehat{p}_{ij,\lambda_m}.$$ 
If $\widehat{p}_{ij}\neq 0$, there is a spillover effect from stock market index $i$ to stock market index  $j$, if $\widehat{p}_{ij}= 0$ there is no spillover effect. 
For instance, the ordered lasso indicates  a spillover effect from \verb|AEX| to \verb|DAX| since the estimated lag length is $\widehat{p}_{13}=1$. In contrast, there is no spillover effect from \verb|AEX| to \verb|DJIA| since $\widehat{p}_{14}=0$. 

In line with the results from the univariate forecast models,
(i) the lag matrix of the hierarchical and ordered lasso is much sparser with overall lower estimated lag lengths than the one of the lasso, and
(ii) American stock market indices have, overall, high own (cfr. diagonal of lag matrix) lag lengths, NIKKEI has overall, a low own lag length.
Furthermore, note that the diagonal of the lag matrix is non-zero (overall, and for all estimators), indicating that each stock market index log-realized variance is influenced by its own past values. 

The lasso-based estimators allow each time series to have its own lag structure in each equation of the VAR, as can be seen from the lag matrices from Figure \ref{Spillover_MULTI}. 
It is important to note that the lag matrix is not a measure of the spillover strengths, but it still allows us to easily distinguish 
``net drivers" and ``net receivers" of stock market index volatility. 
Net drivers influence more other stock market indices than that they are influenced by other indices. 
Net receivers are more influenced by other stock market indices than that they influence other indices.  
We focus on the results of the ordered lasso since it provides the best in-sample fit (i.e BIC= 628.57 versus 908.04 for the hierarchical lasso, and 1065.37 for the lasso).
For some stock market indices, like \verb|FTSE|, the own passed lags are the main driving factor (cfr. near zero FTSE column in lag matrix ordered lasso), others like \verb|AEX| or \verb|EUROSTOXX| are influenced by several other stock market indices. 
Similarly, some stock market indices, like \verb|NIKKEI|,  do not drive any of the other stock market indices' volatility (cfr. zero \verb|NIKKEI| row in lag matrix ordered lasso). 
Overall, we find that American stock market indices are net drivers, the European stock market indices are net receivers, and  \verb|NIKKEI| is a pure receiver of other stock market indices' volatility.
These findings are in line with previous research, see e.g. \citeauthor{Diebold15} (\citeyear{Diebold15}, Chapter 4) and references therein.

\section{Forecast Accuracy \label{ForecastResults}}
We use (i) four estimators: the lasso, hierarchical lasso, ordered lasso, and least squares; (ii) two forecast  combination approaches: forecast combination versus no forecast combination; and (iii) two forecast models: univariate and multivariate.  We compare the forecast performance along these three dimensions  resulting in $16=4\times2\times2$ forecast methods. 

Regarding the forecast combined least squares,  (AR or VAR) models are estimated by least squares, each having lag order $m$, for $1\leq m\leq p_{\text{LS}}$.  The final forecast combines these forecasts by giving the highest weight to the model with the lowest BIC, see equation \eqref{FCeq}. An advantage of the lasso estimators compared to the least squares is that the former are computable if the number of parameters to be estimated exceeds the number of time points, whereas the latter is not.  For this reason, the maximum possible lag length $p_{\text{LS}}$ estimable by least squares might be lower than $p$ depending on the model (AR or VAR) and the number of time points under consideration. For instance, if the number of time points is $\lfloor T/2 \rfloor=98$ (i.e. our smallest estimation window), the maximum possible lag length for the VAR estimable by least squares is $p=9$ since then the number of parameters $p\times q= 9\times 10=90$ per equation is still below the number of time points.
Forecasts without combination are obtained by selecting the sparsity parameter selected based on the BIC for the lasso estimators, and with the lag length selected by the BIC for the least squares.

To evaluate the out-of-sample forecast accuracy of each  method, we use an expanding window approach. We consider different values of the forecast horizon $h\in \{1,2,3,6\}$. At each time point $t= \lfloor T/2 \rfloor,\ldots,T-h$, we use the forecast methods to estimate the AR (in the univariate case) or the VAR model (in the multivariate case) with maximum lag length $p=36$ months. As such, $h$-step ahead direct forecasts  $\widehat{\text{log RV}}_{t+h}^{(i)}$
 and corresponding forecast errors  $e^{(i)}_{t+h} = \text{log RV}_{t+h}^{(i)} - \widehat{\text{log RV}}_{t+h}^{(i)}$ are obtained 
for each stock market index $1\leq i \leq 10$. 
The forecast performance of each stock market index $i$ is then measured by its Absolute Forecast Error
\begin{equation}
\text{AFE}^{(i)}_{h} = \dfrac{1}{T-h-\lfloor T/2 \rfloor+1} \sum_{t=\lfloor T/2 \rfloor}^{T-h} |e^{(i)}_{t+h}|. \nonumber
\end{equation} 
The overall forecast performance, averaged over the ten stock markets, is measured by the Mean Absolute Forecast Error
\begin{equation}
\text{MAFE}_h = \dfrac{1}{10} \sum_{i=1}^{10} \left(\dfrac{1}{T-h-\lfloor T/2 \rfloor+1} \sum_{t=\lfloor T/2 \rfloor}^{T-h} \dfrac{|e^{(i)}_{t+h}|}{\widehat{\sigma}^{(i)}_{1:t}} \right ), \label{amafe}
\end{equation} 
where we divide the forecast errors by $\widehat{\sigma}^{(i)}_{1:t}$, the standard deviation of the log-realized variance of stock market index $i$ computed over the time points $1,\ldots,t$, to adjust for the fact that the different series have different  volatilities and predictabilities (see e.g. \citealp{Carriero11}).

The AFEs for each stock market index are in Table \ref{UNIV_MAFE} for the univariate forecast models, Table \ref{MULTI_MAFE} for the multivariate forecast models. Table \ref{MMAFE_estimator}  summarizes the results by computing the  average $\text{MAFE}_h$ as shown in equation \eqref{amafe} concentrating on the three dimensions Estimator, Forecast Combination, and Forecast Model. 

	\renewcommand{\baselinestretch}{1.1}
\begin{table}
    \sisetup{round-mode=places}
\caption{\footnotesize Univariate Forecast Models.  \label{UNIV_MAFE}}
\scriptsize
\centering
\begin{tabular}{llS[round-precision = 3]S[round-precision = 3]S[round-precision = 3]S[round-precision = 3]S[round-precision = 3]S[round-precision = 3]S[round-precision = 3]S[round-precision = 3]} \hline\hline
{Stock market index} & {Horizon} &\multicolumn{2}{c}{{Least Squares}} & \multicolumn{2}{c}{{Lasso}} & \multicolumn{2}{c}{{Hierarchical Lasso}} & \multicolumn{2}{c}{{Ordered Lasso}} \\
\hline
 && no FC & FC & no FC & FC & no FC & FC & no FC & FC \\ 
  \hline

AEX & $h=1$ & 0.4602 & 0.4578 & 0.5496 & 0.5458 & 0.5812 & 0.5811 & 0.4565 & 0.4576 \\
&  $h=2$& 0.5671 & 0.5526 & 0.5992 & 0.5991 & 0.6259 & 0.6259 & 0.5542 & 0.5526 \\ 
& $h=3$& 0.5990 & 0.5981 & 0.6311 & 0.6276 & 0.6749 & 0.6747 & 0.6188 & 0.6197 \\ 
&  $h=6$& 0.6699 & 0.6710 & 0.7416 & 0.7425 & 0.7168 & 0.7167 & 0.6902 & 0.6924 \\ \hline

CAC & $h=1$ & 0.4938 & 0.4846 & 0.5503 & 0.5485 & 0.5628 & 0.5627 & 0.4832 & 0.4836 \\
&  $h=2$& 0.5873 & 0.5802 & 0.6080 & 0.6095 & 0.6123 & 0.6123 & 0.5777 & 0.5764 \\
& $h=3$& 0.6335 & 0.6337 & 0.6428 & 0.6416 & 0.6297 & 0.6297 & 0.6065 & 0.6109 \\
&  $h=6$& 0.6569 & 0.6563 & 0.6983 & 0.6987 & 0.6825 & 0.6823 & 0.6520 & 0.6551 \\ \hline

DAX & $h=1$& 0.4768 & 0.4722 & 0.5221 & 0.5227 & 0.5492 & 0.5492 & 0.4683 & 0.4694 \\
&  $h=2$& 0.5314 & 0.5305 & 0.5766 & 0.5772 & 0.5851 & 0.5851 & 0.5377 & 0.5380 \\
& $h=3$& 0.5989 & 0.5982 & 0.6488 & 0.6478 & 0.6207 & 0.6208 & 0.5900 & 0.5961 \\
&  $h=6$& 0.6501 & 0.6501 & 0.6776 & 0.8442 & 0.6811 & 0.6814 & 0.6653 & 0.6618 \\ \hline

DJIA & $h=1$ & 0.5667 & 0.5666 & 0.5911 & 0.5926 & 0.6373 & 0.6373 & 0.5479 & 0.5484 \\
&  $h=2$& 0.6496 & 0.6460 & 0.6704 & 0.6690 & 0.7061 & 0.7061 & 0.6383 & 0.6399 \\ 
& $h=3$& 0.6671 & 0.6650 & 0.6880 & 0.6881 & 0.7245 & 0.7245 & 0.6747 & 0.6690 \\ 
&  $h=6$& 0.7331 & 0.7332 & 0.7848 & 0.7855 & 0.7990 & 0.7991 & 0.7531 & 0.7553 \\ \hline

EUROSTOXX & $h=1$ & 0.4766 & 0.4671 & 0.5140 & 0.5172 & 0.5267 & 0.5267 & 0.4685 & 0.4677 \\
&  $h=2$& 0.5523 & 0.5558 & 0.5758 & 0.5730 & 0.5634 & 0.5634 & 0.5439 & 0.5403 \\ 
& $h=3$& 0.6078 & 0.6084 & 0.5933 & 0.5882 & 0.5946 & 0.5946 & 0.5884 & 0.5875 \\ 
&  $h=6$& 0.6308 & 0.6292 & 0.6584 & 0.9505 & 0.6584 & 0.6583 & 0.6319 & 0.6300 \\ \hline

FTSE & $h=1$ & 0.4627 & 0.4576 & 0.5256 & 0.5288 & 0.5504 & 0.5503 & 0.4615 & 0.4617 \\
&  $h=2$& 0.5652 & 0.5586 & 0.5843 & 0.5865 & 0.5952 & 0.5951 & 0.5508 & 0.5502 \\ 
& $h=3$& 0.6148 & 0.6134 & 0.6281 & 0.6193 & 0.6274 & 0.6274 & 0.6059 & 0.6062 \\ 
&  $h=6$& 0.6769 & 0.6768 & 0.6990 & 0.7023 & 0.6897 & 0.6898 & 0.6667 & 0.6686 \\ \hline

NASDAQ & $h=1$ & 0.5214 & 0.5048 & 0.6590 & 0.6583 & 0.7096 & 0.7096 & 0.4980 & 0.4991 \\
&  $h=2$& 0.6107 & 0.6078 & 0.7508 & 0.7357 & 0.7557 & 0.7557 & 0.6012 & 0.6038 \\
& $h=3$& 0.6371 & 0.6344 & 0.7034 & 0.7048 & 0.7763 & 0.7763 & 0.6403 & 0.6435 \\ 
&  $h=6$& 0.6708 & 0.6716 & 0.8408 & 0.8429 & 0.8321 & 0.8321 & 0.7176 & 0.7234 \\ \hline
 
NIKKEI & $h=1$ & 0.6252 & 0.6321 & 0.7320 & 0.7288 & 0.6586 & 0.6592 & 0.6379 & 0.6367 \\ 
&  $h=2$& 0.7187 & 0.7211 & 0.7128 & 0.7054 & 0.6795 & 0.6795 & 0.6840 & 0.6922 \\ 
& $h=3$& 0.7752 & 0.7781 & 0.7671 & 0.7698 & 0.7266 & 0.7267 & 0.7740 & 0.7661 \\ 
&  $h=6$& 0.7619 & 0.7616 & 0.7219 & 0.7227 & 0.7376 & 0.7376 & 0.7407 & 0.7360 \\ \hline

SMI & $h=1$ & 0.5428 & 0.5274 & 0.5663 & 0.5700 & 0.5964 & 0.5963 & 0.5302 & 0.5288 \\
&  $h=2$& 0.6171 & 0.6174 & 0.6416 & 0.6383 & 0.6328 & 0.6329 & 0.6120 & 0.6011 \\ 
& $h=3$& 0.6455 & 0.6433 & 0.6586 & 0.6600 & 0.6724 & 0.6728 & 0.6328 & 0.6298 \\
&  $h=6$& 0.7110 & 0.7123 & 0.7172 & 0.7776 & 0.7484 & 0.7485 & 0.7107 & 0.7107 \\\hline

SP500 & $h=1$ & 0.5435 & 0.5345 & 0.6016 & 0.6017 & 0.6536 & 0.6536 & 0.5296 & 0.5298 \\
&  $h=2$& 0.6613 & 0.6515 & 0.6987 & 0.6987 & 0.7225 & 0.7225 & 0.6394 & 0.6426 \\ 
& $h=3$& 0.6863 & 0.6846 & 0.7177 & 0.7214 & 0.7336 & 0.7337 & 0.6817 & 0.6809 \\ 
&  $h=6$& 0.7412 & 0.7408 & 0.8504 & 0.8450 & 0.8033 & 0.8032 & 0.7632 & 0.7646 \\ \hline\hline
 \multicolumn{10}{p{16cm}}{{\small Notes: Absolute Forecast Errors for the ten stock market indices, the four forecast horizons, and eight forecast methods. ``no FC": result from the optimal model selection by BIC. ``FC": result from Forecast Combination approach.
The sample used is monthly data from January, 2000, to April, 2016, totaling 196 observations. }}
\end{tabular}
\end{table}
	\renewcommand{\baselinestretch}{1.6}

	\renewcommand{\baselinestretch}{1.1}
\begin{table}
    \sisetup{round-mode=places}
\caption{\footnotesize Multivariate Forecast Models. \label{MULTI_MAFE}}
\scriptsize
\centering
\begin{tabular}{llS[round-precision = 3]S[round-precision = 3]S[round-precision = 3]S[round-precision = 3]S[round-precision = 3]S[round-precision = 3]S[round-precision = 3]S[round-precision = 3]} \hline\hline
{Stock market index} & {Horizon} &\multicolumn{2}{c}{{Least Squares}} & \multicolumn{2}{c}{{Lasso}} & \multicolumn{2}{c}{{Hierarchical Lasso}} & \multicolumn{2}{c}{{Ordered Lasso}} \\
 && no FC & FC & no FC & FC & no FC & FC & no FC & FC \\ 
  \hline

AEX & $h=1$ & 0.4719 & 0.4719 & 0.6244 & 0.6222 & 0.6529 & 0.6511 & 0.4681 & 0.4658 \\
& $h=2$& 0.5907 & 0.5907 & 0.6504 & 0.6490 & 0.6735 & 0.6721 & 0.5630 & 0.5617 \\ 
& $h=3$& 0.6320 & 0.6320 & 0.6367 & 0.6369 & 0.6782 & 0.6774 & 0.5991 & 0.5982 \\
&$h=6$ & 0.7004 & 0.7004 & 0.6901 & 0.6901 & 0.7091 & 0.7095 & 0.6784 & 0.6784 \\ \hline

CAC & $h=1$ & 0.5039 & 0.5039 & 0.5973 & 0.5955 & 0.6198 & 0.6182 & 0.4877 & 0.4866 \\
& $h=2$& 0.6265 & 0.6265 & 0.6197 & 0.6183 & 0.6347 & 0.6337 & 0.5793 & 0.5792 \\
& $h=3$& 0.6659 & 0.6659 & 0.6047 & 0.6052 & 0.6369 & 0.6363 & 0.5981 & 0.5974 \\
&$h=6$ & 0.7391 & 0.7391 & 0.6591 & 0.6595 & 0.6727 & 0.6732 & 0.6427 & 0.6427 \\ \hline
 
DAX & $h=1$ & 0.4878 & 0.4878 & 0.6062 & 0.6043 & 0.6246 & 0.6234 & 0.4776 & 0.4767 \\
& $h=2$& 0.5986 & 0.5986 & 0.6212 & 0.6203 & 0.6342 & 0.6334 & 0.5334 & 0.5326 \\
& $h=3$& 0.6284 & 0.6284 & 0.6146 & 0.6149 & 0.6380 & 0.6373 & 0.5860 & 0.5849 \\
&$h=6$ & 0.6767 & 0.6767 & 0.6698 & 0.6701 & 0.6732 & 0.6736 & 0.6519 & 0.6519 \\  \hline

DJIA & $h=1$ & 0.5731 & 0.5731 & 0.6878 & 0.6850 & 0.7151 & 0.7130 & 0.5340 & 0.5331 \\
& $h=2$& 0.6649 & 0.6649 & 0.7063 & 0.7048 & 0.7316 & 0.7303 & 0.6238 & 0.6232 \\
& $h=3$& 0.6448 & 0.6448 & 0.6867 & 0.6872 & 0.7354 & 0.7342 & 0.6656 & 0.6652 \\ 
&$h=6$ & 0.7242 & 0.7242 & 0.7581 & 0.7580 & 0.7868 & 0.7865 & 0.7627 & 0.7630 \\ \hline

EUROSTOXX & $h=1$ & 0.4822 & 0.4822 & 0.5691 & 0.5673 & 0.5903 & 0.5889 & 0.4751 & 0.4746 \\
& $h=2$& 0.6067 & 0.6067 & 0.5882 & 0.5876 & 0.5998 & 0.5987 & 0.5496 & 0.5506 \\ 
& $h=3$& 0.6675 & 0.6675 & 0.5747 & 0.5751 & 0.6025 & 0.6016 & 0.5748 & 0.5746 \\ 
&$h=6$ & 0.6946 & 0.6946 & 0.6214 & 0.6217 & 0.6326 & 0.6332 & 0.6019 & 0.6020 \\ \hline
 
FTSE & $h=1$ & 0.4628 & 0.4628 & 0.5610 & 0.5586 & 0.5969 & 0.5940 & 0.4667 & 0.4648 \\ 
& $h=2$& 0.6279 & 0.6279 & 0.5908 & 0.5897 & 0.6138 & 0.6120 & 0.5464 & 0.5462 \\ 
& $h=3$& 0.6475 & 0.6475 & 0.6115 & 0.6117 & 0.6291 & 0.6284 & 0.5980 & 0.5978 \\ 
&$h=6$ & 0.7276 & 0.7276 & 0.6578 & 0.6577 & 0.6794 & 0.6799 & 0.6657 & 0.6662 \\ \hline

NASDAQ & $h=1$ & 0.5255 & 0.5255 & 0.7307 & 0.7273 & 0.7404 & 0.7392 & 0.5343 & 0.5299 \\
& $h=2$& 0.6284 & 0.6284 & 0.7255 & 0.7236 & 0.7536 & 0.7527 & 0.6449 & 0.6432 \\
& $h=3$& 0.6127 & 0.6127 & 0.7192 & 0.7194 & 0.7552 & 0.7546 & 0.6726 & 0.6726 \\ 
&$h=6$ & 0.6560 & 0.6560 & 0.7683 & 0.7685 & 0.7832 & 0.7830 & 0.7482 & 0.7482 \\ \hline

NIKKEI & $h=1$ & 0.6726 & 0.6726 & 0.7442 & 0.7437 & 0.7390 & 0.7382 & 0.6992 & 0.6926 \\ 
& $h=2$& 0.7509 & 0.7509 & 0.7216 & 0.7217 & 0.7170 & 0.7163 & 0.7311 & 0.7320 \\
& $h=3$& 0.7829 & 0.7829 & 0.7166 & 0.7167 & 0.7245 & 0.7243 & 0.7385 & 0.7393 \\ 
&$h=6$& 0.8868 & 0.8868 & 0.7257 & 0.7257 & 0.7283 & 0.7283 & 0.7231 & 0.7232 \\  \hline

SMI & $h=1$ & 0.5433 & 0.5433 & 0.6519 & 0.6501 & 0.6701 & 0.6681 & 0.5247 & 0.5281 \\ 
& $h=2$& 0.6609 & 0.6609 & 0.6543 & 0.6530 & 0.6831 & 0.6816 & 0.6136 & 0.6131 \\ 
& $h=3$& 0.7014 & 0.7014 & 0.6582 & 0.6587 & 0.6921 & 0.6916 & 0.6445 & 0.6440 \\ 
&$h=6$ & 0.7514 & 0.7514 & 0.7159 & 0.7158 & 0.7266 & 0.7263 & 0.7164 & 0.7167 \\ \hline

SP500 & $h=1$ & 0.5472 & 0.5472 & 0.6915 & 0.6913 & 0.7198 & 0.7179 & 0.5371 & 0.5362 \\
& $h=2$ & 0.6751 & 0.6751 & 0.7245 & 0.7233 & 0.7423 & 0.7414 & 0.6499 & 0.6496 \\ 
& $h=3$ & 0.6778 & 0.6778 & 0.7034 & 0.7039 & 0.7438 & 0.7429 & 0.6843 & 0.6841 \\ 
&$h=6$ & 0.7388 & 0.7388 & 0.7571 & 0.7573 & 0.7901 & 0.7897 & 0.7685 & 0.7685 \\ \hline\hline
  \multicolumn{10}{p{16cm}}{ {\small Notes: Absolute Forecast Errors for the ten stock market indices, the four forecast horizons, and eight forecast methods. ``no FC": result from the optimal model selection by BIC. ``FC": result from Forecast Combination approach.
The sample used is monthly data from January, 2000, to April, 2016, totaling 196 observations. }}
\end{tabular}
\end{table}
	\renewcommand{\baselinestretch}{1.6}

\paragraph{Estimators} 
Table \ref{MMAFE_estimator} (top panel) shows that for all horizons, the ordered lasso attains the lowest, and thus best, MAFE. The least squares performs second best. Among the lasso-based estimators, only the ordered lasso performs better than the least squares. Hence, the largest improvements in forecast accuracy can be obtained by accounting for the dynamic nature of the AR model and imposing monotonicity constraints.

\paragraph{Forecast combination versus no forecast combination}
Table \ref{MMAFE_estimator} (bottom left panel) shows that  the forecast combined approaches produce slightly lower MAFEs than the non-forecast combined ones, and this for all horizons. Hence, the gains of protecting against a wrongly chosen sparsity parameter are, in the majority of considered cases, small.

\paragraph{Univariate versus multivariate forecast models}
Table \ref{MMAFE_estimator} (bottom right panel) shows that for the short forecast horizons $h=1$ and $h=2$, the univariate forecast models outperform the multivariate forecast models.
For $h=3$, the MAFE is about equally good for both. For the longer forecast horizon $h=6$, the multivariate forecast models outperform the univariate ones.
In the long-run, the stock market indices move together, as can be seen from Figure \ref{LogRV}. While it becomes more difficult to obtain accurate forecasts for longer horizons (i.e. MAFEs increase for longer forecast horizons), incorporating spillover effects between the different stock market indices helps to improve forecast accuracy for longer forecast horizons.

	\renewcommand{\baselinestretch}{1}
\begin{table}[t]
\caption{Mean Absolute Forecast Errors ($\text{MAFE}_h$) for the different methods. \label{MMAFE_estimator}}
\centering
\begin{tabular}{lllccccccc}
  \hline\hline
      & \multicolumn{9}{c}{\bf{Estimators}} \\
    \cline{3-10}
Horizon &&& Least Squares && Lasso && Hierarchical Lasso && Ordered Lasso \\ 
  \hline
$h=1$ &&& 0.520 && 0.613 && 0.634 && 0.514  \\ 
$h=2$ &&& 0.624 && 0.650 && 0.663 && 0.599  \\
$h=3$ &&& 0.656 && 0.660 && 0.681 && 0.639  \\
$h=6$ &&& 0.710 && 0.734 && 0.727 && 0.698 \\
\hline
 &&&  &&  &&  &&  \\
Horizon  &&& \multicolumn{3}{c}{\bf{Forecast Combination}} &&  \multicolumn{3}{c}{\bf{Forecast Model}}\\
    \cline{4-6}     \cline{8-10}
 &&& No && Yes && Univariate && Multivariate \\  
  \hline
$h=1$ &&& 0.571 && 0.570  && 0.552 && 0.590  \\ 
$h=2$ &&& 0.634 && 0.633 && 0.622 && 0.646 \\
$h=3$ &&& 0.659 && 0.659  && 0.658 && 0.660 \\
$h=6$ &&& 0.720 && 0.714  && 0.722 && 0.712 \\
 \hline\hline
\multicolumn{10}{p{16cm}}{ {\small Notes: Regarding Estimators: we report the average over the univariate and multivariate forecast models, the forecast combined and non-forecast combined methods. Regarding Forecast Combination: we report the  average over the four estimators, the univariate and multivariate forecast models. Regarding Forecast Model: we report the average over the four estimators, the forecast combined and non-forecast combined methods.
The total sample is monthly data from January, 2000, to April, 2016, totaling 196 observations. }}
\end{tabular}
\end{table}
	\renewcommand{\baselinestretch}{1.6}

\bigskip

In sum, the MAFE measures indicate that (i) the ordered lasso is the best performing estimator; 
(ii) forecast combinations slightly improve the forecast performance of their non-forecast combined alternatives, and (iii) multivariate forecast models only outperform the univariate ones for longer forecast horizons. We further elaborate on these findings by 
inspecting the AFEs for each stock market index separately, see Table \ref{UNIV_MAFE} and \ref{MULTI_MAFE}, and by
computing the Model Confidence Set of \cite{Hansen11}, see Table \ref{MCS}.

\bigskip

\paragraph{AFEs for each stock market index} The ordered lasso also  performs best when looking at the AFEs of each stock market index separately. For the univariate forecast models, it gives the best AFE in 24 out of 40 cases (60\%), thereby clearly outperforming the least squares (best in 35\% of the cases), the hierarchical lasso (best in 5\%) and the lasso (best in 0\%).
For the multivariate forecast models, the ordered lasso gives the best AFE in 25 out of 40 cases (63\%), and performs more than twice as good as the second best which is the least squares (i.e. best in 25\% of the cases).

In the majority of cases, the forecast combinations perform as good as their non-forecast combined alternatives. In some cases, their advantage  is outspoken. In particular,  the forecast combined least squares for the univariate forecast model gives the lowest AFE in 70\% of the cases versus 30\% for its non-forecast combined alternative.
For the multivariate forecast model, the forecast combined ordered lasso attains the best AFE in 78\% of the cases, 
the forecast combined hierarchical lasso in 85\% of the cases. 

Finally, comparing the best univariate to the best multivariate forecasts for each stock market index separately reveals that 
the univariate forecast models attain the best AFE for 8 out of 10 indices, both for horizon $h=1$ and $h=2$. In contrast, for $h=3$ and $h=6$, the multivariate forecast models attain the best AFE for respectively 8 out of 10 and 7 out of 10 cases. Hence, similar conclusions hold as for the MAFEs measures that are averaged over the ten stock market indices.

\bigskip

\paragraph{Model Confidence Set} We use the Model Confidence Set (MCS) procedure to separate the best forecast methods with equal predictive ability from the others. Using the \verb|MCSprocedure| function in \verb|R|, we obtain a MCS that contains the best method with 75\% confidence level. We use the \verb|TR| test statistic computed from 5000 bootstrapped samples, see \cite{Catania14}. The MCS for each forecast horizon is given in Table \ref{MCS}.

For $h=1$,  the MCS includes three (out of the 16) forecast methods: the forecast combined and non-forecast combined ordered lasso, and the forecast combined least squares, all for the univariate forecast model. Hence, univariate forecast models perform best for short horizons.
For $h=2$, the MCS includes five forecast methods. The forecast combined and non-forecast combined ordered lasso for the multivariate forecast model are added to the MCS of horizon $h=1$.
For $h=3$, the MCS includes six forecast methods. The non-forecast combined least squares for the univariate forecast model is added to the MCS of horizon $h=2$.
Finally, for $h=6$, the MCS includes eight forecast models. The forecast combined and non-forecast combined lasso for the multivariate forecast model are added to the MCS of horizon $h=3$. 

	\renewcommand{\baselinestretch}{1.1}
\begin{table}
\small
\caption{Model Confidence set over the 16 forecast methods for the different forecast horizons. \label{MCS}}
 \resizebox{0.78\textwidth}{!}{\begin{minipage}{\textwidth}
 \centering
\begin{tabular}{lcccccccccccc} \hline\hline
&& \multicolumn{4}{c}{{Univariate Forecast Models}} &&& \multicolumn{4}{c}{{Multivariate Forecast Models}} & \\
\cline{3-6} \cline{9-12}
Horizon&& least & lasso & hierarchical & ordered  &&& least & lasso & hierarchical  & ordered & Set size\\ 
&& squares &  & lasso & lasso  &&& squares &  & lasso  & lasso & \\ \hline
$h=1$ && FC & & & no FC; FC &&& & & & & 3\\
$h=2$ && FC & & & no FC; FC &&& & & & no FC; FC & 5\\
$h=3$ && no FC; FC & & & no FC; FC &&& & & & no FC; FC & 6\\
$h=6$ && no FC; FC & & & no FC; FC &&& & no FC; FC & & no FC; FC & 8\\ \hline\hline
\multicolumn{13}{p{21cm}}{\small {Notes: The label ``no FC" (``FC") indicates that the non-Forecast Combined (Forecast Combined) approach of the corresponding estimator and forecast model is included in the MCS. The last column indicates the number of forecast methods in the MCS.
The total sample is monthly data from January, 2000, to April, 2016, totaling 196 observations. }}
\end{tabular}
\end{minipage} }
\end{table}
	\renewcommand{\baselinestretch}{1.6}

In sum, the MCS procedure confirms the excellent performance of the ordered lasso. As the forecast horizon increases, it becomes more difficult to obtain more accurate forecasts, and more forecast methods are included in the MCS.
The MCS contains more multivariate forecast models for longer horizons than for shorter horizons.

\section{Robustness checks \label{robcheck}}
We investigate the robustness of our findings in Section \ref{ForecastResults} to 
(i) the choice of forecast combination weights,
(ii) the length of the sparsity grid $L$,
(iii) the maximum lag length $p$ of the AR/VAR models, and
(iv) the use of rolling versus expanding window forecast approach.

For all robustness checks, similar overall findings to the ones from Section \ref{ForecastResults} are that
(i) the ordered lasso is the best performing estimator,
(ii) forecast combinations perform slightly better  than their non-forecast combined alternatives,
(iii) univariate forecast models perform best for short horizons; multivariate forecast models for longer horizons. 
Below we focus on the new findings of each robustness check. Detailed results are available from the authors upon request.

\paragraph{Forecast combination weights}
As alternative to the BIC-based forecast combination weights of equation \eqref{BICweight}, we take:
(i) equal weights, and 
(ii) MSE-based weights, with weights proportional to the inverse of the individual (in-sample) Mean Squared Estimation Errors (MSE). The highest weight is given to the model with the smallest MSE. 

The ordered lasso retains the best overall forecast performance, regardless of the choice of forecast combination weights. It performs about equally good for all forecast combination weights. The other estimators are, in contrast, more sensitive to an appropriate choice of forecast combination weights. BIC-based weights give the best forecast performance. 

\paragraph{Length of sparsity grid}
In Section \ref{ForecastResults}, we report the results for the length of the sparsity grid $L=20$  since our numerical experiments show that this choice gives the lowest overall MAFEs. The main findings remain the same when varying $L$. The relative performance of the multivariate forecast models to the univariate ones is most influenced by the length of the sparsity grid. 
For a shorter sparsity grid $L=10$, the multivariate forecast models perform much better than the univariate ones. For $h=6$, for instance, their average MAFE is 0.75 compared to 0.92 for the univariate ones, a reduction of 22\%. Since our sparsity grid is logarithmic spaced, a shorter sparsity grid means there are more parsimonious than non-parsimonious models to choose from. Especially the more parsimonious VAR models give more accurate log-realized variance forecasts.
For a longer sparsity grid $L=p=36$, multivariate forecast models perform relatively worse. Now there are more non-parsimonious than parsimonious models to choose from. The non-parsimonious VAR models give, on average, less accurate log-realized variance forecasts.

\paragraph{Maximum lag length} 
We both shorten the maximum lag length from $p=36$ to $p=24$ and extend it to $p=48$. 
The results of all forecast methods are very stable when varying $p$. Only  the performance of the lasso deteriorates when using a larger maximum lag length. The lasso tends to pick up some higher order lags, the more so when $p$ increases. This introduces more noise in the forecast performance and, hence, leads to lower overall AFEs.  Hence, the lasso should only be used as a forecast method for models with lower maximum lag lengths. The hierarchical lasso and ordered lasso, in contrast, do not suffer from this as they encourage low maximal lag lengths in their estimation procedure.

\paragraph{Rolling versus expanding window.}
We replace the expanding window with a rolling window of size $S=\lfloor T/2 \rfloor=98$.
Results are very similar. Only when further decreasing the window size $S$, the performance of the least squares model starts to suffer.  The least squares model becomes imprecise once a lot of parameters need to be estimated relative to the sample size, i.e. $S$. Lasso-based estimators, in contrast, perform regularization to deal with this overparametrization issue, and, as a consequence, retain their good performance. Hence, if few time points are available to make forecasts, the use of lasso-based estimators compared to the least squares becomes more interesting.\\

\section{Discussion \label{conclusion}}
We use lasso-based forecast combination methods to obtain log-realized variance forecasts for ten different stock market indices. Consistently across stock market indices, our results indicate that the ordered lasso is the best performing estimator. The forecast combined ordered lasso slightly improves the forecast accuracy of the non-forecast combined ordered lasso by hedging against the risk of selecting a wrong sparsity parameter. While the ordered lasso for the univariate forecast model performs best for short forecast horizons, the ordered lasso for the multivariate forecast model performs best for longer forecast horizons.

When evaluating the forecast performance of the different forecast methods in Section \ref{ForecastResults}, we compare the predicted log-realized variances to the actual ones over the period February 2008-April 2016.
The Global Financial Crisis, i.e. February 2008-February 2009, falls into this period. We also compute the MAFE for this subperiod only and investigate whether the relative forecast performance of our forecast methods is different during this subperiod. As expected, the MAFEs of all forecast methods are higher compared to the ones from Table \ref{MMAFE_estimator} since it is more difficult to obtain reliable log-realized variance forecasts during this more turbulent period. Nevertheless, the forecast combined ordered lasso remains the best performing forecast method. Multivariate forecast models still have an advantage over the univariate ones for the longer forecast horizons.

One of the objectives of this paper is to compare multivariate with univariate forecast models. To this end, we opted to work with monthly data because of the non-overlapping trading days between the ten stock market indices. Since realized variance forecast are often made at the daily frequency as well, we also compared the forecast performance of the univariate forecast models to the  Heterogeneous AutoRegressive (HAR) model (see e.g. \citealp{Corsi09}) when using daily data. 
In line with the results of Section \ref{ForecastResults}, we find the forecast combined ordered lasso to be the best performing forecast method.
The forecast combined ordered lasso is competitive to the HAR model but it does not  outperform it. 


\bigskip

\noindent
{\bf Acknowledgments.} The authors gratefully acknowledge financial support from the FWO (Research Foundation-Flanders, contract number 12M8217N).




\newpage
\noindent
 
\bibliographystyle{asa}
\bibliography{bibfile}
  
\end{document}